\documentclass[journal=jacsat,manuscript=article,layout=twocolumn]{achemso}
\setkeys{acs}{keywords = true}
\setkeys{acs}{maxauthors = 0} 
\setkeys{acs}{articletitle = true}

\usepackage{chemformula} 
\usepackage[T1]{fontenc} 

\author{Huiyuan Man}
\email{huiyuan653@gmail.com}
\affiliation[a]
{Geballe Laboratory for Advanced Materials, Stanford University, Stanford, CA 94305, USA}
\alsoaffiliation[b]
{Stanford Nano Shared Facilities, Stanford University, Stanford, CA 94305, USA}

\author{Yusuke Iguchi}
\affiliation[a]
{Geballe Laboratory for Advanced Materials, Stanford University, Stanford, CA 94305, USA}
\alsoaffiliation[c]
{Stanford Institute for Materials and Energy Sciences, SLAC National Accelerator Laboratory, Menlo Park, California 94025, USA}

\author{Jin-Ke Bao}
\affiliation[d]
{Materials Science Division, Argonne National Laboratory, Argonne, Illinois 60439, USA}

\author{Duck Young Chung}
\affiliation[d]
{Materials Science Division, Argonne National Laboratory, Argonne, Illinois 60439, USA}

\author{Mercouri G. Kanatzidis}
\affiliation[d]
{Materials Science Division, Argonne National Laboratory, Argonne, Illinois 60439, USA}
\alsoaffiliation[e]
{Department of Chemistry, Northwestern University, Evanston, Illinois 60208, USA}

\title{In-situ local imaging of ferromagnetism and superconductivity in RbEuFe$_4$As$_4$}

\keywords{Ferromagnetic superconductor, superfluid density, magnetic fluctuation, magnetic domain, scanning SQUID microscopy}

\begin{document}

\begin{abstract}

The coexistence of superconductivity and ferromagnetism is an intrinsically interesting research focus in condensed matter physics but the study is limited by low superconducting ($T_c$) and magnetic ($T_m$) transition temperatures in related materials. Here, we used a scanning superconducting quantum interference device to image the in-situ diamagnetic and ferromagnetic responses of RbEuFe$_4$As$_4$ with high $T_c$ and $T_m$. We observed significant suppression of superfluid density in vicinity of the magnetic phase transition, signifying fluctuation-enhanced magnetic scatterings between Eu spins and Fe 3$d$ conduction electrons. Intriguingly, we observed multiple ferromagnetic domains which should be absent in an ideal magnetic helical phase. The formation of these domains demonstrates a weak $c$-axis ferromagnetic component probably arising from Eu spin-canting effect, indicative of possible superconductivity-driven domain Meissner and domain vortex-antivortex phases as revealed in EuFe$_2$(As$_{0.79}$P$_{0.21}$)$_2$. Our observations highlight RbEuFe$_4$As$_4$ is a unique system which includes multiple interplay channels between superconductivity and ferromagnetism.
 
\end{abstract}


Despite the general coexistence of spin-singlet superconductivity and uniform ferromagnetism is not energy-favorable, it had been realized long ago by Anderson and Suhl that a non-uniform magnetic phase could coexist with $s$-wave superconductivity as the ordering wave length of this non-uniform magnetic structure is shorter than the superconducting coherence depth\cite{Anderson_Suhl_1959}. Such unusual coexistence of superconductivity and magnetism has indeed been discovered in the ternary rare-earth compounds such as ErRh$_4$B$_4$, whereas the phase transition temperatures in these materials are relatively low which severely impedes more detailed investigations of the internal structures and intermediate phases of both orders\cite{ERB_1977,review_1985}. The landscape has changed dramatically with the discovery of iron-based superconductors\cite{Fe_based}. Besides the competition and coexistence with antiferromagnetic fluctuations\cite{ThFeAsN_muSR}, superconductivity in iron-based superconductors has also been found to directly coexist with static magnetic order. For example, multiple measurements of Fe(Te,Se) reveal that the bulk superconductivity can coexist with the ferromagnetism in the surface layer\cite{FeTeSe_ARPES,FeTeSe_NVcenter,FeTeSe_Kerr,FeTeSe_transport,FeTe_TI_interface}.

Eu-containing 122-type Fe-based superconductors represent another typical instance of the coexistence of superconductivity and ferromagnetism\cite{(EuK)Fe2As2,EuFe2(AsP)2}. In this superconducting system, the superconducting phase in the FeAs planes develops at relatively higher transition temperatures and a long-range ferromagnetic order in the Eu layers emerges at temperatures below $T_c$\cite{EuFe2(AsP)2,Eu(FeRh)2As2}. A recent magnetic force microscopic measurement of EuFe$_2$(As$_{0.79}$P$_{0.21}$)$_2$ crystal shows the interaction between the superconductivity of Fe 3$d$ electrons and the ferromagnetism from Eu 4$f$ spins introduces remarkable internal spatial structures of both orders in nanoscales\cite{EuFe2(AsP)2_DomainVortex}. In deep superconducting state, a short-period domain Meissner state as well as a nanoscale domain vortex-antivortex state gradually develop as the temperature crosses the magnetic transition temperature $T_m$, featuring emergence of new physics as two antagonistic orders come across\cite{EuFe2(AsP)2_DomainVortex}.

Eu-containing 1144-type superconductor (AEuFe$_4$As$_4$, A = Rb, Cs) provides another ideal platform to study the coexistence and interplay between the magnetic order and superconductivity\cite{AEuFe4As4_Iyo}. This 1144 family adopts a tetragonal lattice structure with Eu layer, two FeAs layers, and Rb/Cs alternatively stacking, leading to a long $c$ axis. $T_{c}$ in this system could reach up to $\sim$ 37 K\cite{AEuFe4As4_Iyo,AEuFe4As4_Cao1,AEuFe4As4_Cao2,RbEuFe4As4_Bao}. Unlike the Eu-containing 122-type family\cite{EuFe2(AsP)2_Euorder,EuFe2(AsP)2_Cao}, the magnetic moments of Eu in 1144 family lie in \textit{ab} plane\cite{RbEuFe4As4_Euorder}. In the meanwhile, the out-of-plane magnetic stiffness in 1144 family is found much smaller than 122 family due to the larger spacial separation between Eu blocks\cite{RbEuFe4As4_Euorder}. As a consequence, the dominant interplay mechanism between Eu spins and Fe 3$d$ electrons in 1144 family is the exchange interaction rather than the electromagnetic interaction as revealed in EuFe$_2$(As$_{1-x}$P$_{x}$)$_2$\cite{EuFe2(AsP)2_Euorder,EuFe2(AsP)2_Cao}. A theoretical analysis based on a Ginzburg-Landau approach suggests that a helical magnetic ground state modulates along the $c$ axis with a period of four times of lattice parameters\cite{RbEuFe4As4_theory_Devizorova}. This non-uniform magnetic phase is predicted to have a superconducting origin\cite{RbEuFe4As4_theory_Devizorova}, highlighting 1144 family potentially harbors interacting physics of superconductivity and magnetism.

\begin{figure*}[t]
\centering
\includegraphics[clip,width=5in]{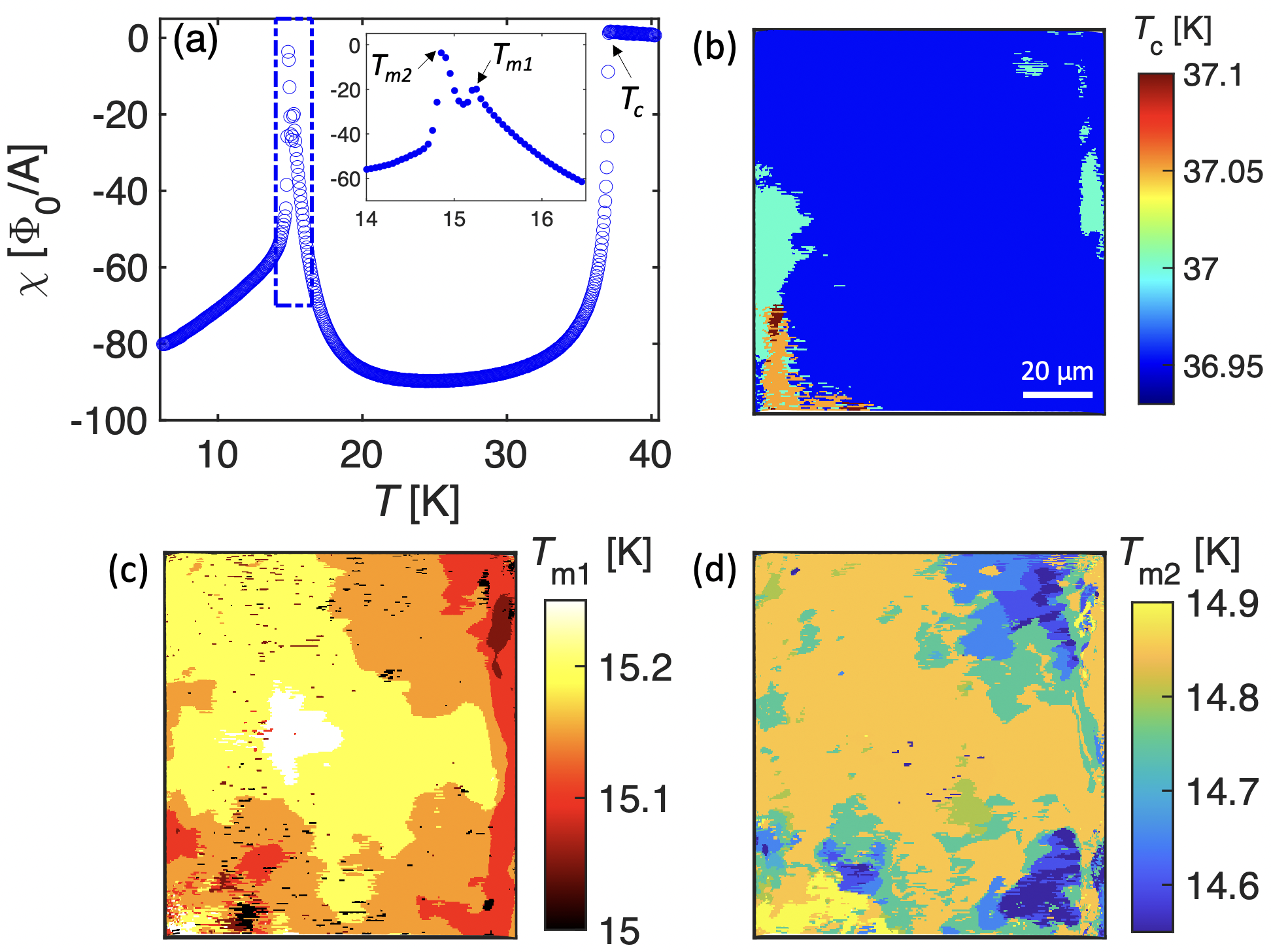}
\caption{(a) The temperature dependence of \textbf{ac} susceptibility measured at 500 Hz of the position specified in SI Figure S1(a). The inset shows the zooming-in profile of the \textbf{ac} susceptibility near the double-peak structure. Maps of (b) local superconducting transition temperature $T_{c}$, local magnetic transition temperatures (c) $T_{m1}$ and (d) $T_{m2}$.
}
\label{SUSC_T}
\end{figure*}

RbEuFe$_4$As$_4$ is one of the 1144 family with $T_c\sim$ 37 K\cite{AEuFe4As4_Cao1,RbEuFe4As4_Bao}. Below $T_m$ $\sim$ 15 K, a helical order develops in the Eu-spin subsystem\cite{AEuFe4As4_Cao1,RbEuFe4As4_Bao}, consistent with the theoretical predictions\cite{RbEuFe4As4_theory_Devizorova}. A few recent experimental studies of this system by high-energy inelastic neutron scattering and infrared spectroscopy indicate rather weak coupling between superconducting and magnetic orders\cite{RbEuFe4As4_Euorder,RbEuFe4As4_vorticesmoveMFM}. However, it is well known that in ferromagnetic systems, magnetic domains generally exist to lower the total free energy. Furthermore, the sample growth process probably introduces some sort of inhomogeneity. These factors make it difficult for the spatial averaged measurements to resolve the intrinsic interaction between magnetism and superconductivity hidden by the nano-/micro-scale magnetic or electronic inhomogeneity\cite{RbEuFe4As4_Euorder,RbEuFe4As4_vorticesmoveMFM}. Here, in order to study the interplay between the superconductivity and magnetic order in RbEuFe$_4$As$_4$, we imaged in-situ diamagnetic and ferromagnetic responses with micrometer scale spatial resolution using scanning superconducting quantum interference device (SQUID) microscopy. The micrometer-level spatial structures of superconductivity and magnetic order were clearly resolved by our scanning microscopy. We found that the superconductivity in RbEuFe$_4$As$_4$ is homogeneous via spatial \textbf{ac} susceptibility imaging near $T_c$. Surprisingly, we observed a dramatic suppression of superfluid density near $T_m$, and the formation of ferromagnetic domains which should be absent in an ideal spin helical phase. These observations indicate multiple interplay channels between superconductivity and magnetism, highlighting a unique coexisting phase in RbEuFe$_4$As$_4$.

\textbf{Sample and technique details.}  RbEuFe$_4$As$_4$ crystal was synthesized through the self-flux method\cite{RbEuFe4As4_Bao}. The cleaved sample surface corresponds to the \textit{ab} plane as illustrated in supporting information (SI) Figure S1(a). To achieve a stable and uniform temperature environment, the sample was glued on a copper block and cooled down in exchange gas. The experiments were performed with the scanning SQUID microscopy, specifically designed to capture the magnetic response along the $c$-axis of RbEuFe$_4$As$_4$ crystal. SQUID is a loop of superconducting Josephson junction (pickup loop) sensitive to the magnetic flux, enabling us to image weak local magnetic fields from the RbEuFe$_4$As$_4$ crystal with sub-micrometer spatial resolution. The local magnetic susceptibility can be measured by integrating a one-turn field coil into the SQUID sensor to apply a localized magnetic field. The inner diameter of pickup loop and field coil is 0.8 $\mu$m and 3 $\mu$m, respectively. The scan height is $\sim\:$1 $\mu$m. The layout of the pickup-loop/field-coil geometry for the SQUID susceptometer is displayed in the SI Figure S1(b). The \textbf{dc} magnetic response is measured in magnetometry mode with unit $\Phi_0$, and the \textbf{ac} magnetic susceptibility is measured in susceptometry mode with the unit $\Phi_0$/A. Here $\Phi_0$ is the quantum magnetic flux $h/2e$. $h$ is the Planck constant and $e$ is the elementary charge. The scanned area of $\sim100$ $\mu$m $\times$ 100 $\mu$m is labelled in the SI Figure S1(a). We conducted susceptibility measurements at various positions and temperatures, relative to the spacing between the sample surface and the SQUID sensor. Fitting the susceptibility data in relation to this spacing allows us to solve the London function and extract the penetration depth (more details in SI)\cite{FitLambda_JohnKirtley}. This method enabled us to determine the temperature-dependent London penetration depth $\lambda$ at specific positions and obtain spatial imaging of $\lambda$.

\begin{figure*}[t]
\includegraphics[clip,width=7.2in]{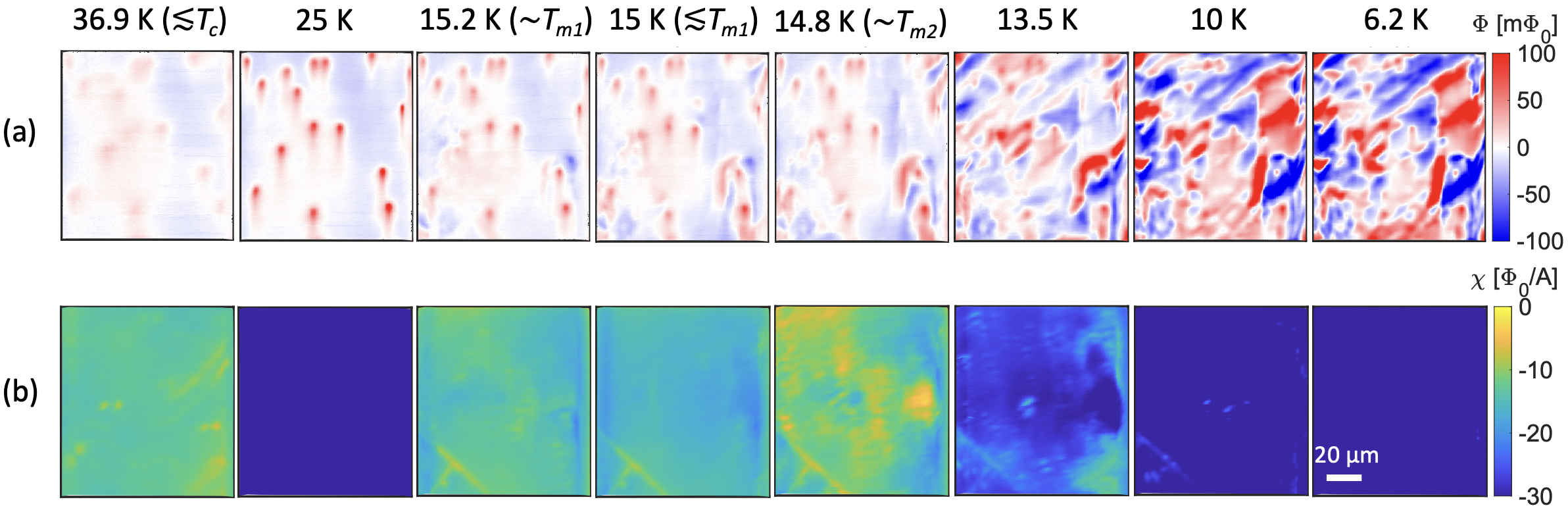}
\caption{Temperature dependence of (a) magnetometry scans and (b) susceptometry scans from 36.9 K to 6.2 K. The data were taken at a number of temperatures during cooling down process. Vortices and magnetic domains are clearly resolved in the magnetometry scans. The inhomogeneity of \textbf{ac} susceptibility is observed around $T_c$ and $T_m$ in the susceptometry scans.
}
\label{MagSuscMap}
\end{figure*}

\textbf{Superconductivity and double magnetic transitions.} As displayed in SI Figure S1(a), the cleaved surface (\textit{ab} plane) of RbEuFe$_4$As$_4$ crystal is shiny and flat, accompanying with some cleaved steps. Figure \ref{SUSC_T}a shows the temperature dependence of \textbf{ac} susceptibility at 500 Hz with SQUID sensor in contact with the sample surface at the position specified in SI Figure S1(a). The susceptibility curve shows sharp superconducting diamagnetism, determining $T_{c}\sim37.05$ K. Here $T_{c}$ is defined as the temperature where the susceptibility drops by 1\% of the total decrease. The spatial mapping of local superconducting transition temperature $T_{c}$ is depicted in Figure \ref{SUSC_T}b. $T_{c}$ in most scanned areas is 36.95 K. There are small areas showing $T_{c}$ of 37.0 K or slightly higher. In general, superconductivity in the whole scanned area is spatially homogeneous, indicating high sample quality.

With cooling below $T_{c}$, the material experiences a magnetic phase transition featured by the pronounced suppression of the diamagnetic signal\cite{RbEuFe4As4_pressure,RbEuFe4As4_muon} as shown in Figure \ref{SUSC_T}a. Intriguingly, the magnetic susceptibility curve $\chi(T)$ reveals a double-peak structure at $T_{m1}$ = 15.25 K and $T_{m2}$ = 14.85 K. To see whether this special structure is intrinsic or not, we obtained the spatial dependence of $T_{m1}$ and $T_{m2}$ with our local scanning capability. As shown in Figures \ref{SUSC_T}c,d, the relevant spatial imaging resolves a few macroscopic areas  with slightly different $T_{m1}$ and $T_{m2}$. Within each macroscopic area, $T_{m1}$ and $T_{m2}$ are very spatially homogeneous. In combination with the spatially homogeneous $T_c$, we conclude that this double-peak structure on \textbf{ac} $\chi(T)$ is an intrinsic property of the crystal. A Ginzburg-Landau analysis of the free energy of RbEuFe$_4$As$_4$ predicts a superconductivity-driven helical magnetic order in the coexisting phase due to the dominant exchange interaction and small $c$-axis magnetic stiffness\cite{RbEuFe4As4_theory_Devizorova}. The susceptibility peak at $T_{m1}$ corresponds to the transition of this helical order. We will show below that our further magnetometry imaging measurement reveals weak $c$-axis ferromagnetism probably arising from Eu spin-canting effect in the helical phase. The susceptibility peak at $T_{m2}$ may result from this spin canting-induced $c$-axis ferromagnetism.

\textbf{Temperature dependent imaging of magnetization and susceptibility.} We took in-situ magnetometry and susceptometry scans at temperatures below $T_c$, illustrated in Figure \ref{MagSuscMap} (more details in SI, Video S1). As shown in Figure \ref{MagSuscMap}a, slightly below $T_c$, the magnetometry scan reveals a few isolated Abrikosov vortices from the weak background field $\sim$ 0.03 Gauss. The magnetic field distribution of vortices extends to the micrometer scale, primarily due to the dimensions of the pickup loop and the scanning height employed in scanning SQUID microscopy\cite{Vortex_John}. When cooling through $T_c$, vortices created by the weak background field at $T_c$ tend to be pinned at the material defects due to the pinning potential\cite{PinningVortex}. The weak "tail" feature of the measured vortices in Figure \ref{MagSuscMap}(a) are due to the magnetic flux penetration through and around the shield in the area of the pickup loop leads\cite{sub_um_SSM_JK}. Using the SQUID's point spread function, we calculated the magnetic fields, which exhibits very similar magnetic structures to those seen in the magnetometry scans (for details, see Supporting Note 5). The vortices do not move as further decreasing temperature (more details in SI Figure S3 and S4). Across $T_{m}$, a few magnetic domains gradually coagulate in the corner of the imaging. As cooling down to the base temperature, magnetic domains with positive and negative magnetization gradually fill the whole scanned surface. The domain width reaches a few tens of $\mu$m at 6.2 K. In strong contrast, the susceptometry scans are relatively spatial homogeneous at all temperatures except near $T_c$ and $T_m$.

The emergence of magnetic domains below $T_m$ is unexpected. As demonstrated by neutron scattering experiment, the magnetic ground state in RbEuFe$_4$As$_4$ crystal is a $c$-axis modulated helical order\cite{RbEuFe4As4_Euorder}. In this state, the Eu spins are constricted in \textit{ab} plane by the magnetic crystalline anisotropy. As cooling down to the base temperature, such a magnetic ground state is expected to impose zero spontaneous magnetization along $c$ axis as well as in \textit{ab} plane. Consequently, no discernible features are anticipated in the magnetometry scans below $T_m$. One hypothesis to reconcile our observation of large ferromagnetic domains and the helical ground state is that besides the primary spin helical order, there exists a weak canting-induced ferromagnetic component along $c$ axis (See Supporting Information, Note 4). As we discussed on Figure \ref{SUSC_T}, the double-peak structure of \textbf{ac} susceptibility may hint the existence of such canting-induced ferromagnetism. Actually, the Eu spin-canting effect has been commonly observed in many Eu-based magnetic materials like EuFe$_2$(As$_{1-x}$P$_x$)$_2$\cite{EuFe2(AsP)2_Euorder}, Eu(Fe$_{1-x}$Ir$_x$)$_2$As$_2$\cite{Paramanik_canting}, EuFe$_2$P$_2$\cite{EuFe2P2_canting}, and Eu(Fe$_{1-x}$Co$_x$)$_2$As$_2$\cite{EuFe2As2_canting}. Previous isothermal magnetization measurements of RbEuFe$_4$As$_4$ have showed a remarkable magnetic hysteresis loop along $c$ axis at 5 K\cite{RbEuFe4As4_anisotropy}. The hysteretic part of the magnetization had been attributed to the vortex pinning effect in superconducting state\cite{RbEuFe4As4_anisotropy,RbEuFe4As4_PNAS}. In our measurement, the spatial resolution of scanning SQUID microscopy is able to distinguish the locations of vortices and magnetic domains. Although we do not exclude the possibility that vortex will contribute to magnetic hysteresis, our detailed temperature dependent magnetometry scans clearly show that the formation of magnetic domains could be an independent or even the main source to the observation of magnetic hysteresis along $c$ axis\cite{RbEuFe4As4_anisotropy}. Here, we would like to point out that the pickup loop of the scanning SQUID sensor primarily captures the out-of-plane component of the magnetic stray field. It may be challenging to entirely rule out the existence of domains with in-plane magnetization.

\begin{figure}[t]
\includegraphics[clip,width=3.5in]{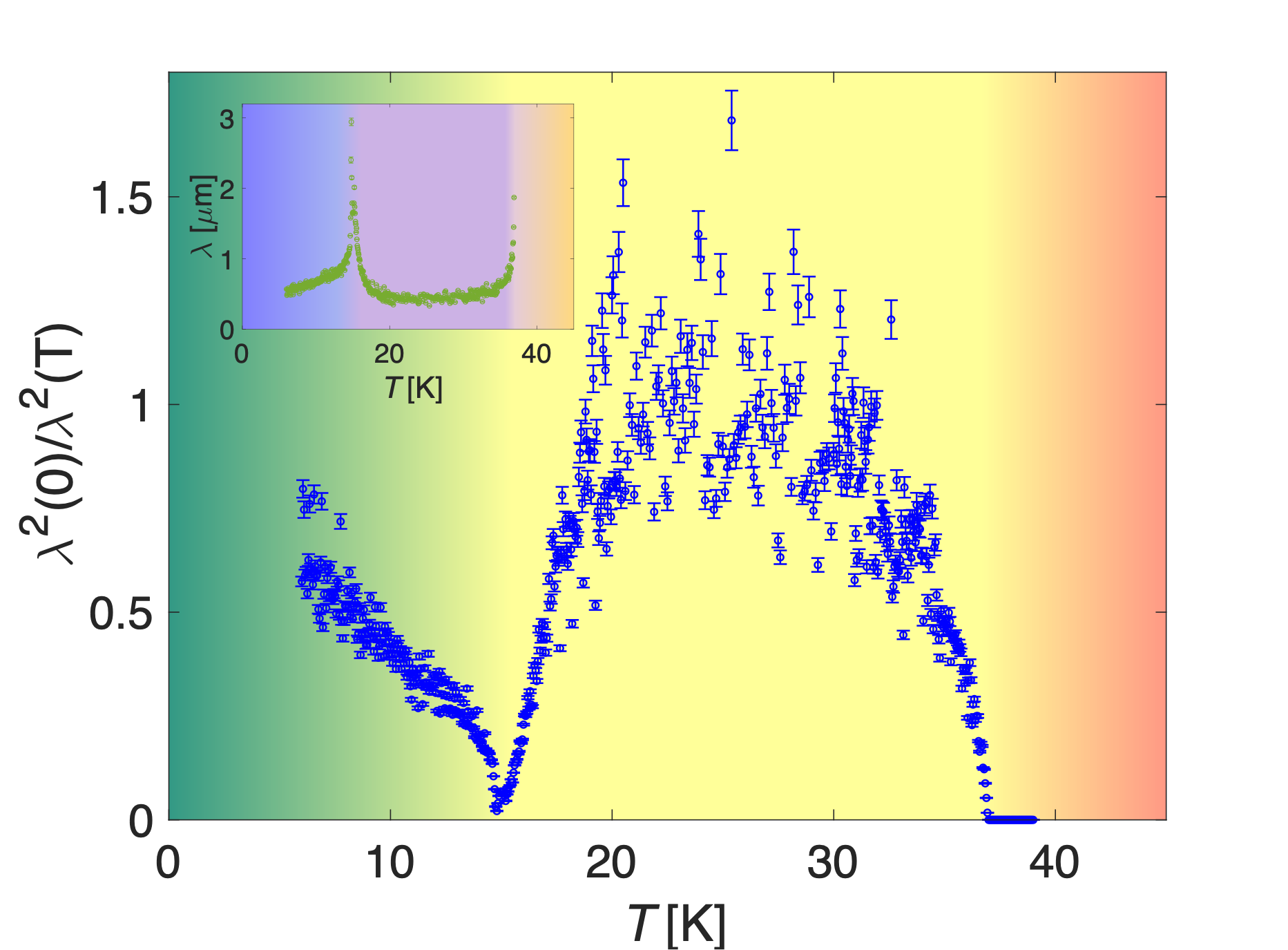}
\caption{Superfluid density $\rho_s\propto\lambda^2(0)/\lambda^2(T)$ as a function of temperature with strong suppression at $\sim T_m$. Inset: Temperature dependent London penetration depth $\lambda$. Error bars are incorporated to illustrate the variability.
}
\label{Lambda}
\end{figure}

\begin{figure*}[t]
\includegraphics[clip,width=5in]{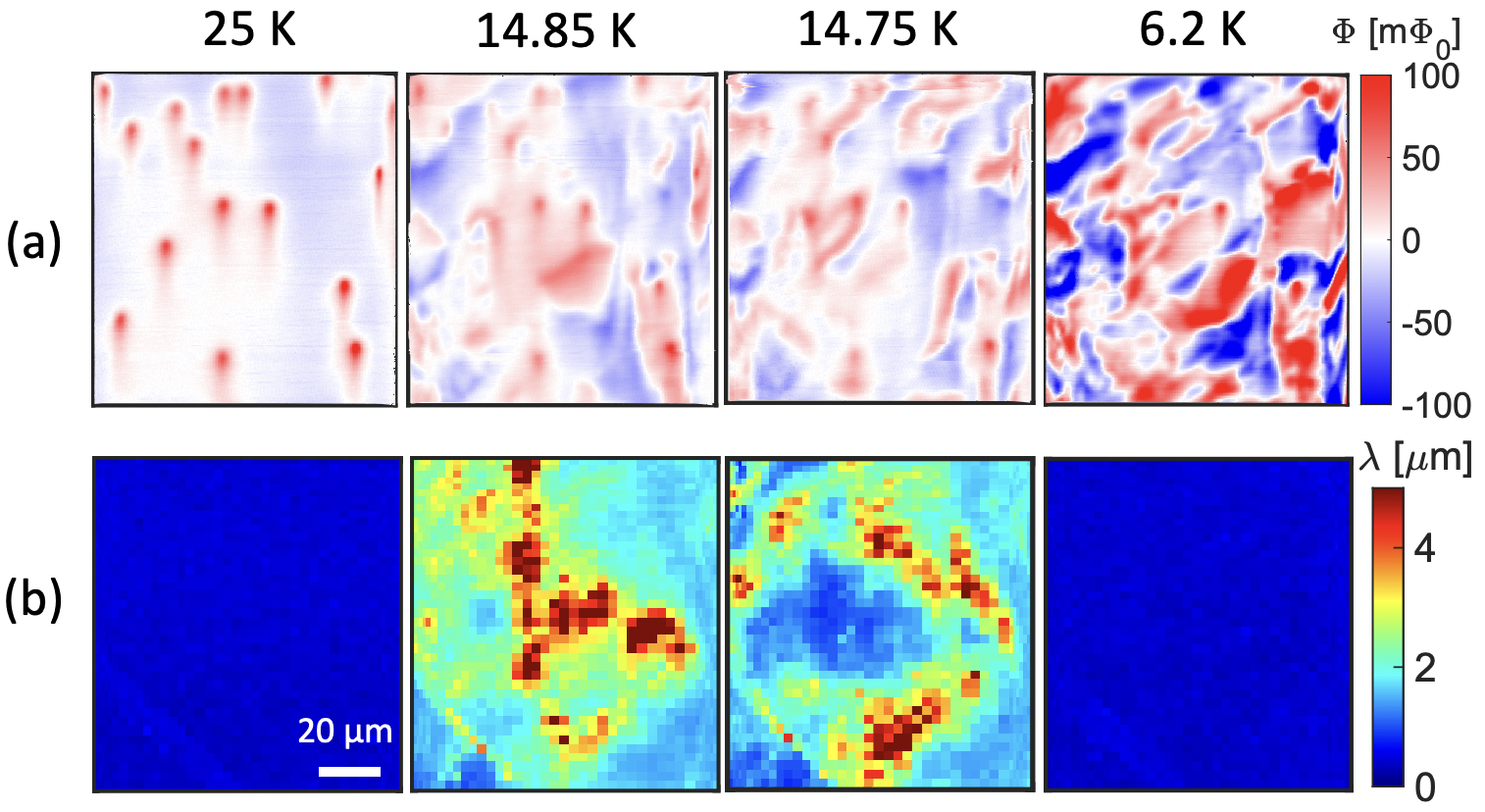}
\caption{The map of (a) magnetometry scans and (b) fitted penetration depth $\lambda$ at 25 K, 14.85 K, 14.75 K, and 6.2 K.
}
\label{Lambdamap}
\end{figure*}

\textbf{Penetration depth and superfluid density.} To further investigate the interplay between the superconductivity and magnetic order in RbEuFe$_4$As$_4$, we illustrated the penetration depth and superfluid density at a specific position labelled by a red cross in SI Figure S1(a). The inset of Figure \ref{Lambda} shows the extracted London penetration depth $\lambda$ which displays a sharp peak at $T_m$. The superfluid density $\rho_s\propto\lambda^2(0)/\lambda^2(T)$ as a function of temperature is shown in Figure \ref{Lambda}. One can see that the superfluid density increases sharply below $T_c$ and then saturates as cooling down to $\sim$ 20 K. Intriguingly, the superfluid density experiences a significant suppression in vicinity of the magnetic phase transition, in consistency with the study of superfluid density by analyzing the temperature dependence of vortex profiles near $T_m$\cite{RbEuFe4As4_SHM}. With further decreasing temperature, the superfluid density starts to rise again but does not recover to the level of 20 K as approaching 6.2 K.

\begin{figure}[t]
\includegraphics[clip,width=3.3in]{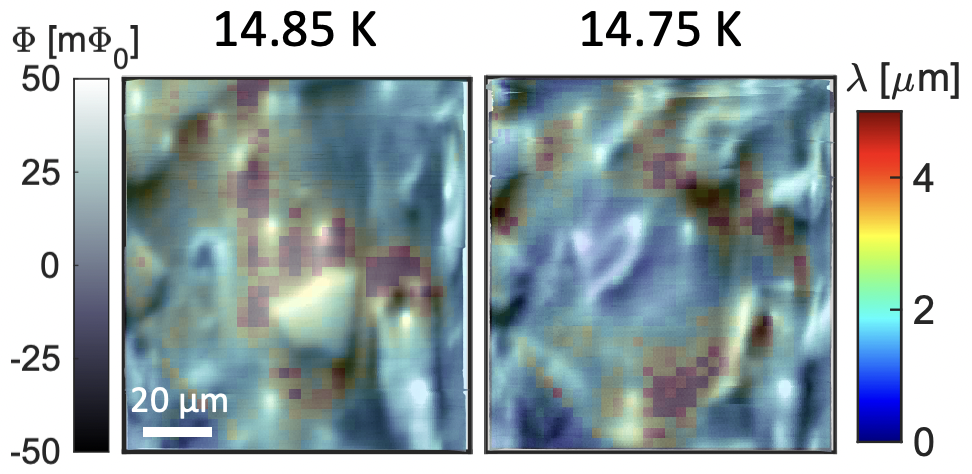}
\caption{The overlapping maps of the penetration depth $\lambda$ and the magnetometry scans at 14.85 K and 14.75 K. The enhancement of $\lambda$ mainly appears in areas where the out-of-plane magnetic domains have not formed yet.
}
\label{Overlap}
\end{figure}

In addition to the temperature dependent London penetration depth and superfluid density at a specific position, we further investigated their spatial mappings in the vicinity of the magnetic transition. As presented in Figure \ref{Lambdamap}a, at 25 K, a temperature in the deep superconducting state but still above $T_m$, the penetration depth $\lambda$ $\propto$ $1$/$\sqrt{\rho_s}$ is small and spatially homogeneous. Nevertheless, at 14.85 K and 14.75 K, the penetration depth mapping becomes spatially inhomogeneous. Meanwhile, small magnetic domains start to develop at the same temperatures, 14.85 K and 14.75 K, near $T_m$ (Figure \ref{Lambdamap}a). At 6.2 K, $\lambda$ and $\rho_s$ become spatially homogeneous again, although the scanned area is full of large magnetic domains.

\textbf{Discussion.} The large suppression of superfluid density $\rho_s$ near $T_m$ is interesting. As we demonstrated in the magnetometry scans, far below $T_m$, a lot of ferromagnetic domains form over the scanned surface. Although macroscopic magnetization exists in these domains, the superfluid density does not experience significant suppression at 6.2 K, distinct from the observations near $T_m$. This sharp contrast indicates that the macroscopic magnetization itself is not the dominant factor to significantly suppress superconductivity. To further emphasize this point, we overlapped magnetometry scans and penetration depth mappings around $T_m$. We show relevant plots in Figure \ref{Overlap}. One can see that the enhancement of $\lambda$ (suppression of $\rho_s$) mainly appears in regions where the out-of-plane magnetic domains have not formed yet. In the areas where out-of-plane magnetic domains develop, the suppression of superfluid density is much weaker. Therefore, the formation of magnetic domains seems to restrain the main factor which significantly suppresses superfluid density near $T_m$.

In the context of the Abriosov-Gorkov theory, paramagnetic impurities usually act as pair-breaking centers to deplete the superfluid density and smear the coherence peaks of density of states in superconductors\cite{Abrikosov_1960,NbN_sc_disordered_2016}. Above $T_m$, the paramagnetic state of Eu sub-lattice imposes a pair-breaking effect to the superconductivity in RbEuFe$_4$As$_4$, whereas the suppression of superfluid density is weak (Figure \ref{Lambda}). As shown by ref. \cite{suppression_sc_theory}, with the system approaching a magnetic phase transition, the magnetic correlation length increases rapidly, which enables a condition that the magnetic correlation length becomes comparable to the superconducting coherence length. In this special regime, magnetic phase fluctuation is significant and Eu spins become strongly correlated. The spin-flip scattering between Eu spins and Fe 3$d$ conduction electrons will be significantly enhanced as well, resulting in a large suppression of superfluid density. We observed a rapid drop and recovery of superfluid density in Figure \ref{Lambda}. This behavior defines a broad temperature range near $T_m$ that the magnetic correlation length is comparable to the superconducting coherence length, suggesting a relatively small superconducting coherence length in RbEuFe$_4$As$_4$\cite{suppression_sc_theory}. It is noteworthy that the superfluid density becomes quite spatially inhomogeneous as it is strongly depleted near $T_m$ (Figure \ref{Lambdamap}). Such a low and spatially inhomogeneous superfluid density around $T_m$ indicates cooperative behaviors of magnetism and superconductivity in vicinity of the magnetic phase transition. Meanwhile, the formation of magnetic domains could restrain the magnetic fluctuation near $T_m$, which weakens the magnetic fluctuation-mediated suppression of superfluid density.

An ideal spin helical order modulated along $c$ axis as proposed in RbEuFe$_4$As$_4$ previously should give rise to zero spontaneous magnetization in all crystallographic orientations\cite{RbEuFe4As4_Euorder}. Our observation of ferromagnetic domains indicates that besides the primary magnetic helical order, a weak $c$-axis ferromagnetic component exists, making RbEuFe$_4$As$_4$ an analog to the ferromagnetic superconductor EuFe$_2$(As$_{0.79}$P$_{0.21}$)$_2$\cite{EuFe2(AsP)2_DomainVortex}. A recent magnetic force microscopic study revealed that the ferromagnetism and superconductivity in EuFe$_2$(As$_{0.79}$P$_{0.21}$)$_2$ coexist in a non-uniform manner\cite{EuFe2(AsP)2_DomainVortex}. Slightly below the ferromagnetic transition temperature $T_{\rm{FM}}$, EuFe$_2$(As$_{0.79}$P$_{0.21}$)$_2$ develops a domain Meissner state. The domain width is reported to be 100 to 200 nm. Further below $T_{\rm{FM}}$, a new domain vortex-antivortex phase emerges and the domain size reaches $\sim$ 350 nm\cite{EuFe2(AsP)2_DomainVortex}. The weak ferromagnetism along $c$ axis in RbEuFe$_4$As$_4$ immediately calls attentions on whether or not similar domain Meissner state and domain vortex-antivortex phase exist in the coexisting phase. From our magnetometry scans above and below $T_m$ (Figures \ref{MagSuscMap}, \ref{Lambdamap}), this is possible. We demonstrated that RbEuFe$_4$As$_4$ crystal develops complicated domain structures over the whole scanned \textit{ab} plane. These magnetic domains may embody more intricate structures in the sub-micrometer scale. Unfortunately, our scanning SQUID microscopy is not able to probe the internal domain structures in the hundreds of nanometer scale due to the limited spatial resolution. Further experiments with better spatial resolution are necessary to clear this issue.

In summary, we performed systematic spatially resolved studies on RbEuFe$_4$As$_4$ to investigate the conexistence and interplay of superconductivity and ferromagnetism. By taking advantage of the scanning SQUID microscopy, we were able to characterize the in-situ magnetic and superconducting responses of RbEuFe$_4$As$_4$ in magnetometry and susceptometry modes. The homogeneous superconductivity with \textit{T$_c$} $\sim$ 37 K was demonstrated by the mapping of local superconducting transition temperature. Interestingly, we observed a double-peak structure from the temperature dependent \textbf{ac} susceptibility. We attributed the susceptibility peak at $T_{m1}$ to the transition of superconductivity-driven helical order, and the other susceptibility peak at $T_{m2}$ to the formation of a weak $c$-axis ferromagnetic component. We showed the temperature dependent and space-resolved London penetration depth and superfluid density of RbEuFe$_4$As$_4$, highlighting that the magnetic fluctuations of Eu moments strongly suppress the superconductivity in vicinity of the magnetic phase transition. The formation of ferromagnetic domains below $T_m$, which is unexpected in an ideal helical spin phase, features RbEuFe$_4$As$_4$ is an analogy to the ferromagnetic superconductor EuFe$_2$(As$_{0.79}$P$_{0.21}$)$_2$\cite{EuFe2(AsP)2_DomainVortex}. Our observations highlight that RbEuFe$_4$As$_4$ is a unique system which embodies multiple channels for the interplay between superconductivity and magnetism like the superconductivity-driven helical order\cite{RbEuFe4As4_theory_Devizorova}, the enhanced scatterings between Eu spins and Fe 3$d$ conduction electrons in vicinity of the magnetic phase transition\cite{suppression_sc_theory}, and the possible domain Meissner and domain vortex-antivortex phases\cite{EuFe2(AsP)2_DomainVortex}.

\begin{acknowledgement}

The authors thank Kathryn A. Moler and Y. Masaki for fruitful discussions. This work was primarily supported by the US Department of Energy, Office of Basic Energy Sciences, Division of Materials Sciences and Engineering under award DE-AC02-76SF00515. In addition, the sample synthesis and crystal growth at Argonne National Laboratory was supported by the U.S. Department of Energy, Office of Science, Basic Energy Sciences, Materials Sciences and Engineering Division. Experiments utilized equipment in the Stanford Nano Shared Facilities, funded by the National Science Foundation under award ECCS-2026822.

\end{acknowledgement}

\begin{suppinfo}

The sample surface, susceptometry and magnetometry scans, fitting method and procedure for the susceptibility as a function of sample-sensor spacing, clarification of the weak $c$-axis magnetization, scanning SQUID's point spread function.

\end{suppinfo}

\bibliography{ref}

\end{document}